\documentclass[english,keywords,amsmath,amssymb,twocolumn]{revtex4}
\usepackage[T1]{fontenc}
\usepackage[latin1]{inputenc}
\usepackage{braket}
\usepackage{babel}%
\usepackage{graphicx}
\usepackage{color}
\usepackage{bm}
\usepackage{longtable}
\usepackage{amsmath}
\usepackage{amsfonts}
\usepackage{dsfont}
\usepackage{amssymb}
\usepackage{hyperref}
\begin{document}
\title{Steering inequality for pairs of particle-number-superselection-rule restricted states}
\author{Asmita Kumari}
\author{Ujjwal Sen}
\affiliation{Harish-Chandra Research Institute, HBNI, Chhatnag Road, Jhunsi, Allahabad 211 019, India}

\begin{abstract}
We consider violations of a Clauser-Horne-Shimony-Holt-type steering inequality for quantum states of systems of indistinguishable particles restricted by a particle-number-superselection rule. We check for violations in non-interacting Bose-Einstein condensate  and N00N states, by using two-copies of the states for bypassing the superselection rule. The superselection rule prevents the states from maximally violating the steering inequality. But the steering inequality violations are higher than Bell inequality violations for the same states. This in particular implies, in certain cases, that  visibilities of the steering inequality violations are higher than the same for Bell inequality violations, for admixtures with white noise. We also found that an increase in the number of particles in the non-interacting condensate states results in a decrease of the violation amount of the steering inequality.


\end{abstract}

\maketitle
\section{Introduction}
Entanglement is a fundamental resource of quantum information processing~\cite{ent}. Entanglement is typically considered between separated systems or different degrees of freedom of the same physical system, for which the tensor-product structure is well-defined. 
For the case of indistinguishable particles, it is therefore natural to consider entanglement between different modes. 


Violation of Bell inequalities forms an important method for detecting entanglement present in quantum states of shared systems~\cite{bell64}. 
%
%
The Bell Clauser-Horne-Shimony-Holt (CHSH) inequality~\cite{chsh69} for two distinguishable systems, each having the option to choose between two dichotomic measurement settings, $A_{1}, A_{2}$ and $B_{1}, B_{2}$, respectively for the two parties, is given by \(|B| \leq 2\), where
\begin{eqnarray}
\label{B}
B=\langle A_{1} \otimes B_{1}\rangle+\langle A_{1} \otimes B_{2}\rangle+\langle A_{2} \otimes B_{1}\rangle-\langle A_{2} \otimes B_{2}\rangle. ~ 
\end{eqnarray}
 The maximum violation of this inequality that can be reached by any quantum state is 
 $2 \sqrt{2}$~\cite{cri}.

A straightforward testing of Bell's inequality may not be possible for systems of indistinguishable particles, since
rotations away from fixed  particle  number bases may not be allowed due to the restriction imposed by a superselection rule on  indistinguishable systems~\cite{wick52, wise03, suhu04a, pagol-hawa1, pagol-hawa2, pagol-hawa3, pagol-hawa4}.
 %
Interestingly,  Heaney, Lee, and Jaksch~\cite{hea10} derived a method for testing Bell inequalities for states of indistinguishable particles even within the superselection-rule restrictions. 
In order to overcome the superselection rule, they use two copies of system for performing rotated measurements. 

Quantum steering, introduced by Schr{\" o}dinger~\cite{sch35}, is a concept that lies in between those of  entanglement and Bell inequality violation~\cite{Qui}. If two observers share an entangled state, then unsteerability implies that there will exist a ``local hidden state'' model of the state. 
Steerability can be tested through the violation of steering inequalities (see e.g.~\cite{reid89,cav09,Ou92,wal11,sch11}).  The violation of steering inequalities, like that of Bell inequalities, indicates the presence of entanglement in the shared state involved.

Recently, Cavalcanti, Foster, Fuwa, and Wiseman~\cite{cav16} have proposed a CHSH-type inequality to  check for  steerability in the  two-party, two-setting (per party), and two-outcome (per measurement setting) scenario. Using the same notations as for the Bell inequality, 
the steering inequality is given by \(S\leq 2\), where
\begin{eqnarray}
\label{S}
\nonumber S =\sqrt{\langle (A_{1}+A_{2})  \otimes B_{1}\rangle^{2}+\langle (A_{1}+A_{2}) \otimes B_{2}\rangle^{2}} \quad \quad \\  +\sqrt{\langle (A_{1}-A_{2}) \otimes B_{1}\rangle^{2}+\langle (A_{1}-A_{2})  \otimes B_{2}\rangle^{2}}. 
\end{eqnarray}
In quantum mechanics, the optimal  violation of this steering inequality is again $2 \sqrt{2}$. 

Our goal in this paper is to consider violation of the steering inequality on quantum systems of indistinguishable particles that are restricted by particle-number superselection rules. We check for violation in non-interacting Bose-Einstein condensate states and ``N00N'' states. We find that compared to Bell-CHSH  inequality violation, the steering inequality violation reaches a higher visibility for admixture with white noise, in certain cases. Just like for the Bell inequality,  maximum violation for the steering inequality is however still less than what is possible without the superselection rule.

The rest of the paper is organized as follows.
In Sec.~II, we set the notations corresponding to the   steering inequality, and propose to use it for two copies of bimodal states. In Sec.~III, we test this inequality for  non-interacting Bose-Einstein-condensate states, and compare their violations with Bell inequality violations for the same states. In Sec.~IV, we consider the same test for the N00N states
Finally in Sec.~V we conclude our findings.

\section{Steering inequality }

Let us consider two systems, which are each split into two spatially non-overlapping modes. Suppose that $N_1$ and  $N_2$ are particle numbers of the first and second systems respectively, and let the composite system state be $\sigma^{N_1+N_2} = \rho^{N_1}_{a b} \otimes \rho^{N_2}_{A B}$. Here $a$ and $A$ are  two modes of the two systems, controlled by say, Alice, and $b$ and $B$ are the two further modes of the two systems, supervised by say, Bob. Let us now assume that Alice performs a joint measurement on her two modes by using the operator, 
\begin{eqnarray}
A(\phi_j) = \sum^{N_1+N_2}_{n_c +m_C =0} \epsilon (n_c , m_C)  (|n_c , m_C\rangle \langle n_c , m_C |)_{c C}
\end{eqnarray}
where $\epsilon (n_c , m_C)$ is a weighting coefficient. Similarly as Alice, Bob makes a joint measurement on his two modes, $b$ and $B$, which is denoted by $ B (\theta_k)$. Here $j , k =1, 2$ denotes the two measurement settings each of Alice and Bob. Measurements by Alice and Bob on their respective modes only allow to perform local particle number measurements. However, in order to perform general measurements, the spatial modes of both the systems are separately allowed to pass through separate beam-splitters. 
For Alice, the beam-splitter transformation is given by
\begin{equation}
\hat{c} = \alpha \hat{a}+\beta \exp(-i \phi_j) \hat{A},            \enspace \hat{C} = \beta \hat{a}-\alpha \exp(-i \phi_j) \hat{A},
\end{equation}
where $\hat{a}$  and $\hat{A}$ are annihilation operators corresponding to the two input modes of Alice's beam-splitter, and   $\hat{c}$  and $\hat{C}$ are annihilation operators for the two output modes of the same. A similar transformation is true on Bob's side:
\begin{eqnarray}
\hat{d} = \alpha \hat{b}+\beta \exp(-i \theta_k) \hat{B},                                     \enspace \hat{D} = \beta \hat{b}-\alpha \exp(-i \theta_k) \hat{B}.
\end{eqnarray}
Here  $\hat{b}$  and $\hat{B}$ are annihilation operators for the two input modes of Bob's beam-splitter and   $\hat{d}$  and $\hat{D}$ are annihilation operators for the two output modes of the same. 
It is noted that each party measures in a particle number basis. The output of the measurement depends on the local angles, $ \phi_j$ and $\theta_k$, of Alice's and Bob's beam-splitter settings respectively. The measurement vector  $ |n_c , m_C\rangle$ associated with the observable $A(\phi_j)$ for the particle numbers  $n_c $ and $ m_C$ in the output modes $c$ and $C$ is given by 
\begin{eqnarray}
 |n_c , m_C\rangle_{c C} = \phantom{ami baRi jabo gaache choRbo peyara aa}\nonumber\\  \frac{(\alpha \hat{a}^{\dagger}+\beta e^{-i \phi_j} \hat{A}^{\dagger})^{n_c}}{\sqrt{n_c !}}  \frac{(\beta \hat{a}-\alpha e^{-i \phi^j} \hat{A})^{m_C}}{\sqrt{m_C !}}|0 , 0\rangle_{a A},
\end{eqnarray}
where $|0 , 0\rangle_{a A}$ is vacuum state corresponding to the modes $a$ and $A$.  We now set the weighting coefficient $\epsilon (n_c , m_c)$ of $A(\phi_j)$ as~\cite{hea10} 
\begin{eqnarray}
\label{mauni-tapas}
 \epsilon (n_c , m_C) = (-1)^{m_C + \frac{(m_C+n_c)(m_C+n_c+1)}{2}}.
\end{eqnarray}
Since in the composite system,  there are $N_1 + N_2$  particles, the number of outcomes of the measurements is 
\begin{eqnarray}
\label{out}
O =\left[ \frac{1}{2}(N_1 + N_2)+1\right](N_1 + N_2+1). 
\end{eqnarray}
The CHSH  Bell inequality  purposed in \cite{hea10} for pairs of states restricted with the  particle-number-superselection rule is given by
\begin{eqnarray}
\label{bp}
\nonumber |B_p|& \equiv &| \langle A(\phi_1) \otimes  B (\theta_1)  \rangle+\langle  A(\phi_1) \otimes  B (\theta_2) \rangle \\  &+& \langle   A(\phi_1) \otimes  B (\theta_2)\rangle-\langle   A(\phi_2) \otimes  B (\theta_2) \rangle | \leq2,
\end{eqnarray}
where
the 
correlation between the observables is defined as
\begin{eqnarray}
\nonumber
 \langle A(\phi_j) \otimes  B (\theta_k)  \rangle  = \sum_{n_c +m_C+n_d+m_D=N_1+N_2} \epsilon (n_c , m_C) \times \\  \epsilon (n_d, m_D)  P( A(\phi_j) \otimes  B (\theta_k)), \quad
\end{eqnarray}
with  $P( A(\phi_j) \otimes  B (\theta_k))$ denoting the joint probability of outcomes of measurement of the local observables  $A(\phi_j) $ and $B (\theta_k)$.
Similar to how  this Bell inequality was formed using local observables $ A(\phi_j) $ and $  B (\theta_k)  $,  we can write the CHSH-type steering inequality 
in Eq.~(\ref{S}) using the same local observables. Each observable on Alice's or Bob's side is bounded: $ \parallel A(\phi_j) \parallel \leq1$,  \(\parallel B (\theta_k)  \parallel \leq 1\).
Therefore, the  steering inequality which bypasses the superselection rule 
can be formulated as 
\begin{eqnarray}
\label{Ss}
\nonumber S_p&\equiv&  \Big(\left\langle (A(\phi_1)+A(\phi_2)) \otimes  B (\theta_1)\right\rangle^{2}\\ \nonumber &+&\left\langle (A(\phi_1)+A(\phi_2)) \otimes B (\theta_2)\right\rangle^{2}\Big)^{1/2}\\ \nonumber &+&\Big(\left\langle (A(\phi_1)-A(\phi_2) )\otimes B (\theta_1)\right\rangle^{2}\\ &+&\left\langle (A(\phi_1)-A(\phi_2))\otimes B (\theta_2)\right\rangle^{2}\Big)^{1/2} \leq2.
\end{eqnarray}
In the two succeeding sections,  we apply this steering inequality to the cases of non-interacting Bose-Einstein condensate states and the ``N00N'' states, known to be useful in precision measurements. In order to compare the violation of the steering inequality with that of Bell inequality for the states, we first recapitulate in each case, the known results for the Bell inequality in~(\ref{bp}).

\section{Non-interacting Bose-Einstein condensate states}
In order to test the steering inequality for non-interacting Bose-Einstein condensate states, we have considered two different cases.\\

\noindent $Case~(i)$:    When both systems share the same number of particles, i.e.,  $N_1 = N_2 = N$.\\

Consider 
the zero-temperature non-interacting Bose-Einstein condensate state with a fixed particle number, symmetrically 
distributed between  two modes: 
\begin{eqnarray}
\label{shrabana-ghanai}
|\psi_N \rangle = \frac{1}{{\sqrt{2}}^N}\sum^N_{n=0} \frac{\sqrt{N!}}{\sqrt{n! (N-n)!}}|n , N-n\rangle.
\end{eqnarray}
If the number of particle of each system is unity ($N = 1$), then $|\psi_N \rangle$ reduces to 
$|\psi_1 \rangle = \frac{1}{\sqrt{2}}(|10\rangle + |01\rangle)$, and the composite system is in the state,
\begin{eqnarray}
|\psi'_1 \rangle = \frac{1}{2}(|10\rangle + |01\rangle)\otimes (|10\rangle + |01\rangle).
\end{eqnarray}
Since the composite system has a total of two particles ($N_1 + N_2 = 2 N =2$), the  number of measurement outcomes is $6$ (see Eq.~(\ref{out})). In case that the beam-splitter is balanced, i.e., when $\alpha = \beta =1/\sqrt{2}$, the six measurement settings  on modes $a$ and $A$ with  weighting coefficients $ \epsilon (n_c , m_C) $ on Alice's side is given in Table I.





\begin{widetext}
\begin{center}

\begin{table}

\begin{tabular}
{|p{1.1 cm} | c| c|}
\hline 
\centering
  $|n m \rangle_{c, C}  $   &   Effective measurement basis for Alice's modes \(a\) and \(A\)    & $  \epsilon (n_c , m_C) $ \\
\hline
\hline
\centering
  $ |0 0 \rangle $   &  $ |0 0 \rangle  $    & $  1 $ \\
\hline
\centering
  $ |1 0 \rangle $ & $ \frac{1}{\sqrt{2}}(|1 0 \rangle + \exp({-i \phi_1})|0 1 \rangle)  $  &  $-1 $ \\
\hline
 \centering
  $ |0 1 \rangle $   & $ \frac{1}{\sqrt{2}}(|1 0 \rangle - \exp({- i \phi_1})|0 1 \rangle)  $ &  $ 1 $\\
\hline
\centering
  $  | 1 1 \rangle $ &  $ \frac{1}{\sqrt{2}}(|2 0 \rangle - \exp({- 2 i  \phi_1})|0 2 \rangle) $   & $ 1 $ \\
\hline
\centering
  $  | 2 0 \rangle $ &  $ \frac{1}{2}(|2 0 \rangle +\sqrt{2}\exp({- i \phi_1})|1 1 \rangle + \exp({- 2i  \phi_1})|0 2 \rangle) $   & $ -1 $ \\
\hline
\centering
  $  | 0 2 \rangle $ &  $ \frac{1}{2}(|2 0 \rangle - \sqrt{2}\exp({- i \phi_1})|1 1 \rangle + \exp({- 2i  \phi_1})|0 2 \rangle) $   & $ -1 $ \\
\hline
\end{tabular}
\caption{[This table is given in Ref.~\cite{hea10}.] When we make a measurement on the modes \(c\) and \(C\) on Alice's side, in the particle number basis, an effective measurement  is carried out on Alice's side in the modes \(a\) and \(A\). For the case when \(N=1\), this effective measurement basis is given in the middle column. The corresponding \(\epsilon(n_c, m_C)\) is given in the last column. 
}
\end{table}
\end{center}
\end{widetext}

The six measurement settings 
on the modes $b$ and $B$ of  Bob's side can be similarly obtained.  Now, using these measurement settings of Alice and Bob for the system state $|\psi'_1 \rangle$, the  steering inequality expression is given by
\begin{eqnarray}
\label{Ss}\nonumber
S^{BEC}_{p,1} &=& \frac{1}{2} \bigg[\bigg((\cos (\theta_1-\phi_1)+\cos (\theta_1-\phi_2)-2\big)^2\\ \nonumber&+&\big(\cos (\theta_2-\phi_1)+\cos (\theta_2-\phi_2)-2)^2\bigg)^{1/2}\\ \nonumber&+&\sqrt{2} \bigg(\sin ^2\left(\frac{\phi_1-\phi_2}{2}\right) (2-\cos (2 \theta_1-\phi_1-\phi_2)\\ &-&\cos (2 \theta_2-\phi_1-\phi_2))\bigg)^{1/2}\bigg].
\end{eqnarray}
The maximum value of $S^{BEC}_{p,1} $ is $\approx 2.79$, and is reached  at $\phi_1 =0, \phi_2 =\pi/2, \theta_1 \approx 3.93$ and $ \theta_2 \approx 2.90$. The quantum mechanical expression for Bell-CHSH expression from (\ref{bp}) for the same quantum state and the same operators is given by
\begin{eqnarray}
\label{Bs}
B^{BEC}_{p,1} &=& \frac{1}{2} (-\cos (\theta_1-\phi_1)-\cos (\theta_1-\phi_2)\\ \nonumber&-&\cos (\theta_2-\phi_1)+\cos (\theta_2-\phi_2)+2).
\end{eqnarray}
 The maximum value of $|B^{BEC}_{p,1}|$ is  $\approx 2.41$, and is reached at $\phi_1 =0, \phi_2 =\pi/2, \theta_1 \approx 3.93$ and $ \theta_2 \approx 2.36$, as obtained in~\cite{hea10}. 
 We therefore have reached at the following two observations for the state under consideration:
\begin{itemize}
\item[1.] Constraints of particle-number-superselection rule on measurement space prevents the maximum violation from reaching $2 \sqrt{2}$ for both Bell as well as  steering inequality.

\item[2.] Optimal violation of  steering inequality 
is 
higher 
than that for the Bell inequality, and therefore the steering inequality violation will have greater visibility than the Bell inequality violation for admixture with white noise, at least in cases where the observables involved are traceless.
\end{itemize}

Admixing with white noise for a state \(|\psi\rangle\) is defined as the creation of the state \(p|\psi\rangle \langle \psi| + (1-p) \rho_w\), where \(p \in [0,1]\) is the admixing probability and \(\rho_w\) is the completely depolarized state of the relevant dimension. The claim made in item 2 above about visibility  holds if \(A_i \otimes B_j\) for  \(i,j=1,2\) and \((A_1 \pm A_2) \otimes B_k\) for \(k=1,2\) are traceless. This is true, e.g., for the case when either \(N_1\) or \(N_2\) is unity. 

\begin{figure}[ht]
	\includegraphics[width=1\linewidth]{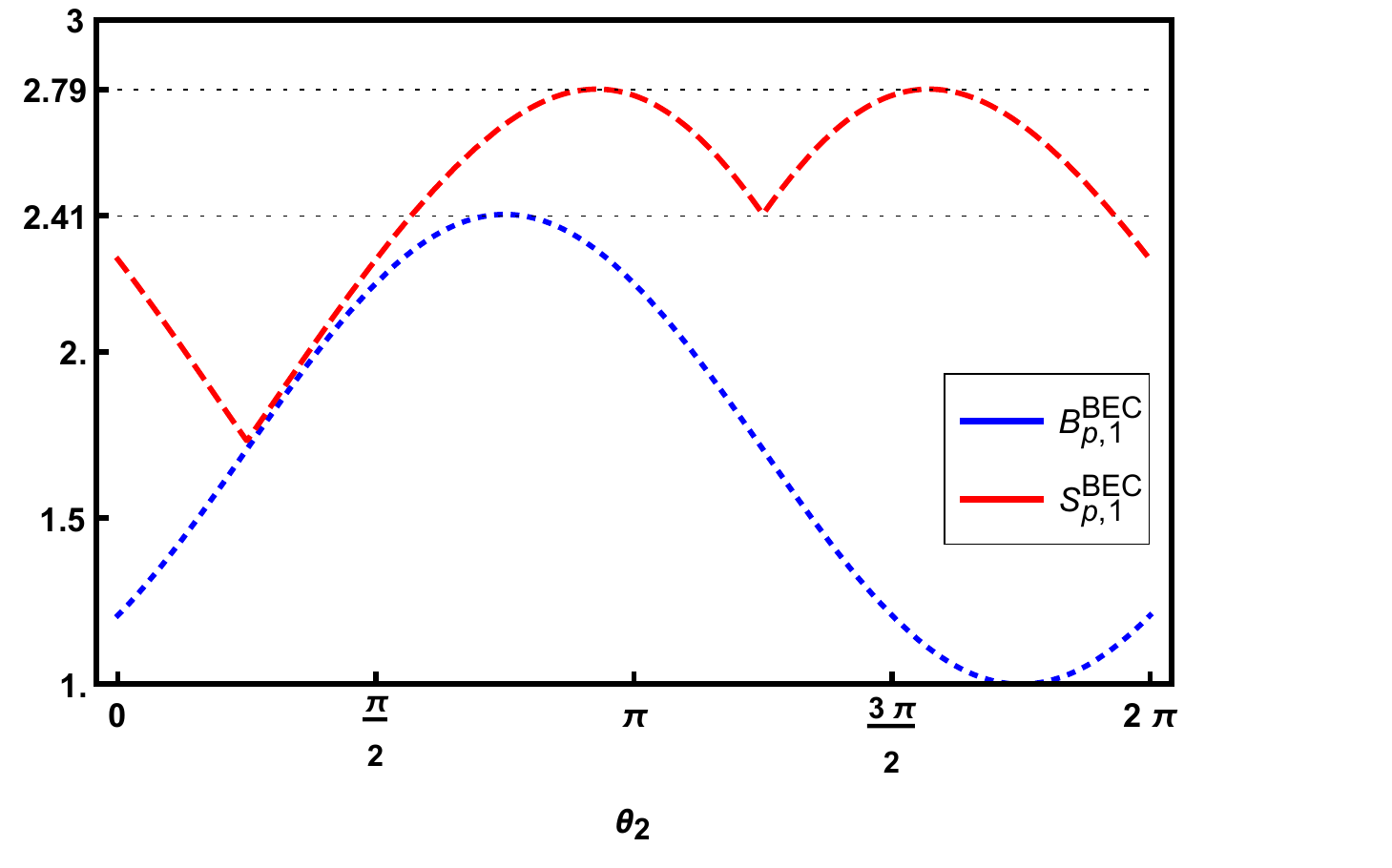}
	\caption{Comparing steering inequality violation with Bell inequality violation. The quantities  $S^{BEC}_{p,1} $ and $|B^{BEC}_{p,1}|$ are plotted with respect to the measurement parameter,  $\theta_2$, at $\phi_1 =0, \phi_2 =\pi/2, \theta_1=3.93$. The state involved is given by two copies of the state given in Eq.~(\ref{shrabana-ghanai}), for \(N=1\). The horizontal axis is in radians, while the vertical one is dimensionless.}
	\label{fig:1}
\end{figure}

 In Fig. 1, we have plotted  $S^{BEC}_{p,1} $ and $|B^{BEC}_{p,1}|$ at $\phi_1 =0, \phi_2 =\pi/2, \theta_1=3.93$ with respect to $\theta_2$. From the figure, we can see that there are two maxima of steering inequality violation in the region \(\theta_2 \in [0,2\pi)\) compared to a single maximum for  violation 
 of Bell inequality. This can potentially be useful in an actual implementation to check for violation of the steering inequality.

Next, let us consider two particles  in each system, i.e., $N_1 = N_2 = N = 2$. Then, $|\psi_N \rangle$ reduces to 
$|\psi_2 \rangle = \frac{1}{2}(|02\rangle +\sqrt{2} |11\rangle+ |20\rangle)$, and the composite state of the two systems is 
\begin{equation}
|\psi'_2 \rangle = \left(\frac{1}{2}(|02\rangle +\sqrt{2} |11\rangle+ |20\rangle)\right)^{\otimes 2}.
\end{equation}
The number of measurement outcomes is now 15 (compare with Eq.~(\ref{out})).
The fifteen measurement settings corresponding to $15 $ effective outcomes of the modes $a$ and $A$ of Alice with  weighting coefficient $ \epsilon$, as given in Table II. 

\begin{widetext}
\begin{center}

\begin{table}
\begin{tabular}{|p{1.1 cm} | c| c|}
\hline 
\centering
  $|n m \rangle_{c, C}  $   &   Effective measurement basis for Alice's modes \(a\) and \(A\)    & $  \epsilon (n_c , m_C)  $ \\
\hline
\hline
\centering
  $ |0 0 \rangle $   &  $ |0 0 \rangle  $    & $  1 $ \\
\hline
\centering
  $ |1 0 \rangle $ & $ \frac{1}{\sqrt{2}}(|1 0 \rangle + \exp({-i \phi_1})|0 1 \rangle)  $  &  $-1 $ \\
\hline
 \centering
  $ |0 1 \rangle $   & $ \frac{1}{\sqrt{2}}(|1 0 \rangle - \exp({- i \phi_1})|0 1 \rangle)  $ &  $ 1 $\\
\hline
\centering
  $  | 1 1 \rangle $ &  $ \frac{1}{\sqrt{2}}(|2 0 \rangle - \exp({- 2i  \phi_1})|0 2 \rangle) $   & $ 1 $ \\
\hline
\centering
  $  | 2 0 \rangle $ &  $ \frac{1}{2}(|2 0 \rangle +\sqrt{2}\exp({- i \phi_1})|1 1 \rangle + \exp({- 2i  \phi_1})|0 2 \rangle) $   & $ -1 $ \\
\hline
\centering
  $  | 0 2 \rangle $ &  $ \frac{1}{2}(|2 0 \rangle - \sqrt{2}\exp({- i \phi_1})|1 1 \rangle + \exp({- 2i  \phi_1})|0 2 \rangle) $   & $ -1 $ \\
\hline
\centering
  $  | 1 2 \rangle $ &  $ \frac{1}{4}(|3 0 \rangle - \sqrt{2}\exp({- i \phi_1})|2 1 \rangle  - \sqrt{2}\exp({- 2i  \phi_1})|1 2 \rangle+ \exp({-3 i  \phi_1})|0 3 \rangle) $   & $ 1 $ \\
\hline
\centering
  $  | 2 1 \rangle $ &  $  \frac{1}{4}(|3 0 \rangle +\sqrt{2}\exp({- i \phi_1})|2 1 \rangle  - \sqrt{2}\exp({- 2i  \phi_1})|1 2 \rangle- \exp({- 3i  \phi_1})|0 3 \rangle) $   & $ -1 $ \\
\hline
\centering
  $  | 0 3 \rangle $ &  $ \frac{1}{4 \sqrt{3}}(|3 0 \rangle - 3\exp({- i \phi_1}) |2 1 \rangle  + 3\exp({- 2i  \phi_1})|1 2 \rangle- \exp({- 3i  \phi_1})|0 3 \rangle) $   & $ -1 $ \\
\hline
\centering
  $  | 3 0 \rangle $ &  $  \frac{1}{4\sqrt{3}}(|3 0 \rangle +3\exp({- i \phi_1})  |2 1 \rangle  + 3\exp({- 2i  \phi_1})  |1 2 \rangle+ \exp({- 3i  \phi_1})|0 3 \rangle) $   & $ 1 $ \\
\hline
\centering
  $  | 2 2 \rangle $ &  $ \frac{1}{8}(|4 0 \rangle -2\exp({- 2i  \phi_1})  |2 2 \rangle + \exp({- 4i  \phi_1})|0 4 \rangle) $   & $ 1 $ \\
\hline
\centering
  $  | 4 0 \rangle $ &  $  \frac{1}{8\sqrt{6}}(|4 0 \rangle +4\exp({- i \phi_1})  |3 1 \rangle  + 6\exp({- 2i  \phi_1})  |2 2 \rangle+ 4\exp({- 3i  \phi_1})  |1 3 \rangle)+\exp({- 4i  \phi_1})|0 4 \rangle) $   & $ 1 $ \\
\hline
\centering
  $  | 0 4 \rangle $ &  $  \frac{1}{8\sqrt{6}}(|4 0 \rangle -4\exp({- i \phi_1})  |3 1 \rangle  + 6\exp({- 2i  \phi_1})  |2 2 \rangle- 4\exp({- 3i  \phi_1})  |1 3 \rangle)+\exp({- 4i  \phi_1})|0 4 \rangle) $   & $ 1 $ \\
\hline
\centering
  $  | 1 3 \rangle $ &  $  \frac{1}{4\sqrt{6}}(|4 0 \rangle -2\exp({- i \phi_1})  |3 1 \rangle + 2\exp({- 3i  \phi_1})  |1 3 \rangle)-\exp({- 4i  \phi_1})|0 4 \rangle) $   & $ -1 $ \\
\hline
\centering
  $  | 3 1 \rangle $ &  $  \frac{1}{4\sqrt{6}}(|4 0 \rangle +2\exp({- i \phi_1})  |3 1 \rangle - 2\exp({- 3i  \phi_1})  |1 3 \rangle)-\exp({- 4i  \phi_1})|0 4 \rangle)  $   & $ -1 $ \\
\hline
\end{tabular}
\caption{The considerations are the same as in Table I, except that we have \(N_1=N_2=2\) here.
}
\end{table}
\end{center}

\end{widetext}

Similarly, the fifteen effective measurement settings  on the modes $b$ and $B$  on Bob's side can be obtained.  Now, using these measurement settings of Alice and Bob for the composite system state $|\psi'_2 \rangle$, the quantum expression of the steering inequality is given by
\begin{eqnarray}
\label{Ss2}\nonumber
S^{BEC}_{p,2} &=& \Big[\left(\sin ^4\left(\frac{\phi_1-\theta_1}{2}\right)-\sin ^4\left(\frac{\phi_2-\theta_1}{2}\right)\right)^2\\  \nonumber &+&\left(\sin ^4\left(\frac{\phi_1-\theta_2}{2}\right)-\sin ^4\left(\frac{\phi_2-\theta_2}{2}\right)\right)^2\Big]^{1/2}\\  \nonumber &+&   \Big[\left(\sin ^4\left(\frac{\phi_1-\theta_1}{2}\right)+\sin ^4\left(\frac{\phi_2-\theta_1}{2}\right)\right)^2\\  &+&\left(\sin ^4\left(\frac{\phi_1-\theta_2}{2}\right)+\sin ^4\left(\frac{\phi_2-\theta_2}{2}\right)\right)^2\Big]^{1/2}.\quad \quad
\end{eqnarray}
The maximum value of  $S^{BEC}_{p,2}$ is  $\approx 2.78$ at $\phi_1 =0, \phi_2 \approx 1.07, \theta_1 \approx 3.93$ and $ \theta_2 \approx 3.00$. For the same state and the same measurement settings, the 
Bell operator (see (\ref{bp})) is has the quantum mechanical average value of
\begin{eqnarray}
\label{Bs2}
B^{BEC}_{p,2} &=& \sin ^4\left(\frac{\phi_1-\theta_1}{2}\right)+\sin ^4\left(\frac{\phi_2-\theta_1}{2}\right) \nonumber\\   &+&\sin ^4\left(\frac{\phi_1-\theta_2}{2}\right)-\sin ^4\left(\frac{\phi_2-\theta_2}{2}\right).
\end{eqnarray}
The maximum value of $|B^{BEC}_{p,2}|$ is $\approx 2.36$ and is reached at $\phi_1 =0, \phi_2 \approx 1.07, \theta_1 \approx 3.68$ and $ \theta_2 \approx 2.60$. 
Comparing with the \(N_1=N_2=1\) case, we find that the quantum violation has decreased with increasing particle number for both the steering as well as the Bell inequality.
%
However, this decrease in violation for the steering inequality is lower than that for the  Bell inequality.
\begin{figure}[ht]
	\includegraphics[width=1\linewidth]{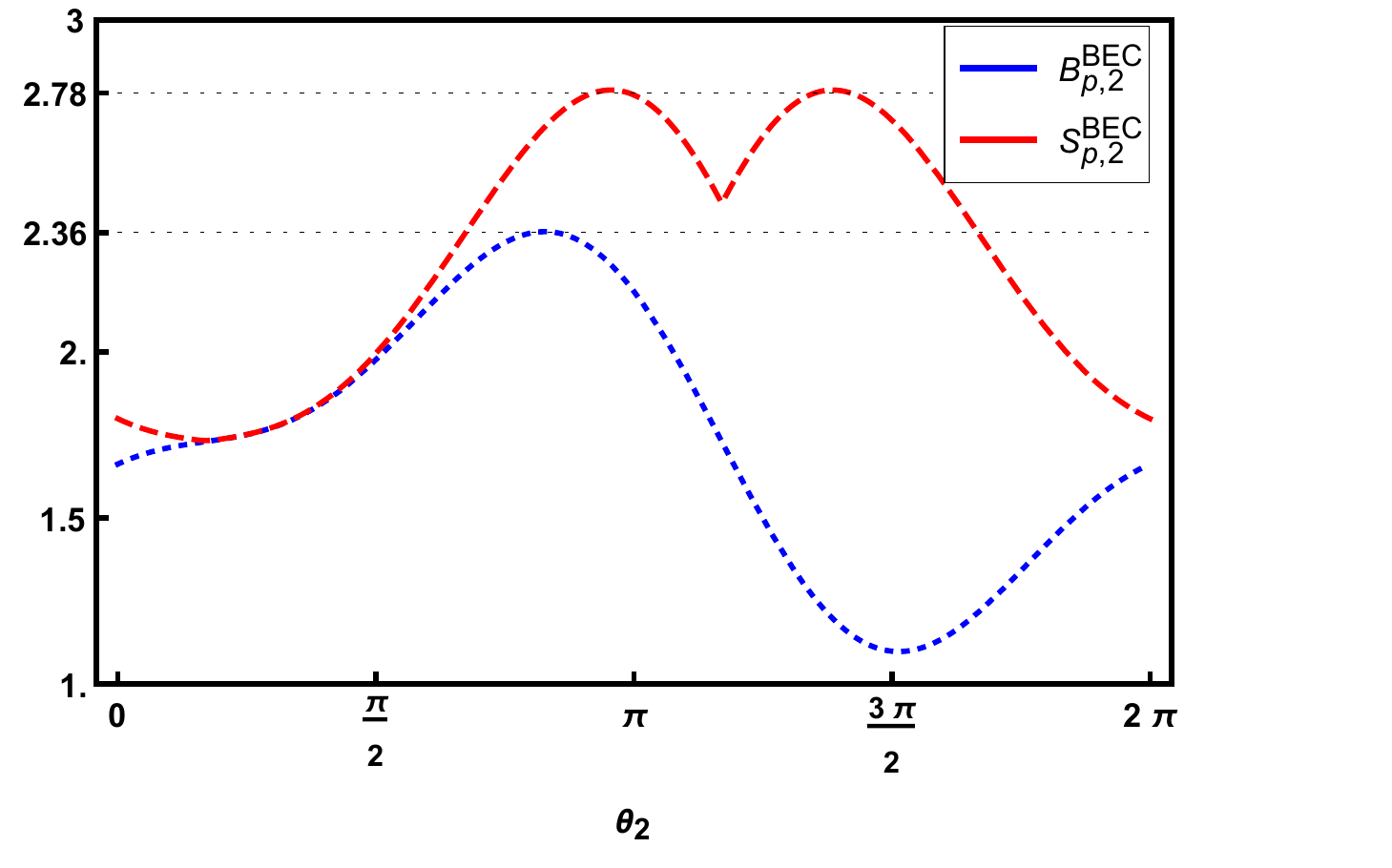}
	\caption{
	The considerations are the same as in Fig. 1, except that \(N_1=N_2=2\) here. Also, the Bell inequality curve is plotted for \(\theta_1=3.68\), while the steering inequality one is plotted for \(\theta_1=3.93\). 
\(\phi_1 =0, \phi_2 =1.07\) for both the curves. 
}
	\label{fig:1}
\end{figure}
 In Fig. 2 we have plotted  $S^{BEC}_{p,2} $ and $|B^{BEC}_{p,2}|$ at  $\phi_1 =0, \phi_2 =1.07$, with respect to $\theta_2$. 
 We have chosen \(\theta_1 = 3.93\) for the former, while \(\theta_1=3.68\) for the latter.
 We see in Fig. 2 that similar to the case when \(N_1=N_2=1\), there are two maxima of the steering inequality violation in the \(\theta_2\)-parameter space, while there is a single maximum for the Bell inequality violation in the same space. \\

\noindent $Case~(ii)$:     When the two systems share different numbers of particles, i.e., $N_1 \neq N_2 $.\\

The steering inequality is now to be checked for the state 
$|\psi_{N_1} \rangle \otimes |\psi_{N_2} \rangle $. In the case when $N_1 = 1$ and $N_2 = 2$, the composite state is
\begin{eqnarray}
|\psi_{1,2} \rangle = \frac{1}{2 \sqrt{2}} (|10\rangle + |01\rangle)  \otimes (|02\rangle +\sqrt{2} |11\rangle+ |20\rangle),
\end{eqnarray}
and number of measurement outcomes is now 10 (compare with  Eq.~(\ref{out})). These can be read off from the 
first $10$ rows 
in  Table II.  If the beam-splitter is balanced, i.e., if 
$\alpha = \beta = 1/\sqrt{2}$, all correlation functions of the steering inequality vanish. 
For unbalanced beam splitters the   correlation functions are not vanishing, but still no violation of the steering inequality is obtained.


\section{N00N states}
Consider the two-mode state, 
%
\begin{eqnarray}
|\{N, m\} \rangle = \frac{1}{ \sqrt{2}} (|N-m,m\rangle + |m, N-m\rangle), 
\end{eqnarray}
and for definiteness, suppose that 
$N=2$ and $m=0$.
Suppose also that the  composite state of Alice and Bob is now 
\begin{eqnarray}
|\psi'_3 \rangle = |\{2, 0\} \rangle^{\otimes 2} = \left(\frac{1}{ \sqrt{2}} (|2,0\rangle + |0, 2\rangle)\right)^{\otimes 2}.
\end{eqnarray}
 For this state, 
 the expression in the steering inequality is given by
\begin{eqnarray}
\label{Sn}\nonumber
S^n_{p} &=&\bigg(\left(\cos ^2(\theta_1-\phi_1)+\cos ^2(\theta_1-\phi_2)\right)^2\\ \nonumber&+&\left(\cos ^2(\theta_2-\phi_1)+\cos ^2(\theta_2-\phi_2)\right)^2\bigg)^{1/2}\\ \nonumber&+&\frac{1}{\sqrt{2}}\bigg(\sin ^2(\phi_1-\phi_2) (2-\cos (4 \theta_1-2 (\phi_1+\phi_2))\\ &-&\cos (4 \theta_2-2 (\phi_1+\phi_2))-2)\bigg)^{1/2}.
\end{eqnarray}
The maximum  violation of the steering inequality is  $\approx 2.79$, which is the same as that obtained for the non-interacting Bose-Einstein condensate state with a single particle in each system. 
The quantum violation of $ 2.79$ is now obtained at $\phi_1 \approx -0.13$, $\phi_2 \approx 0.65$, $\theta_1 \approx 0.26$ and $\theta_2 \approx 0.672$. The Bell expression for the same quantum state and same observables is given by
\begin{eqnarray}
\label{Bn}
B^n_{p} &=&-\sin (\theta_1-\theta_2) \sin (\theta_1+\theta_2-2 \phi_2) \nonumber\\&+&\cos ^2(\theta_1-\phi_1)+\cos ^2(\theta_2-\phi_1).
\end{eqnarray}
The maximum of \(|B^n_p|\) is $ \approx 2.41$,  as obtained for the Bose-Einstein condensate state with a single particle in each system. The quantum violation of $ 2.41$ is  now obtained at $\phi_1 \approx -0.13$, $\phi_2 \approx 0.65$, $\theta_1 \approx 0.26$ and $\theta_2 \approx -0.52$.   
\begin{figure}[ht]
	\includegraphics[width=1\linewidth]{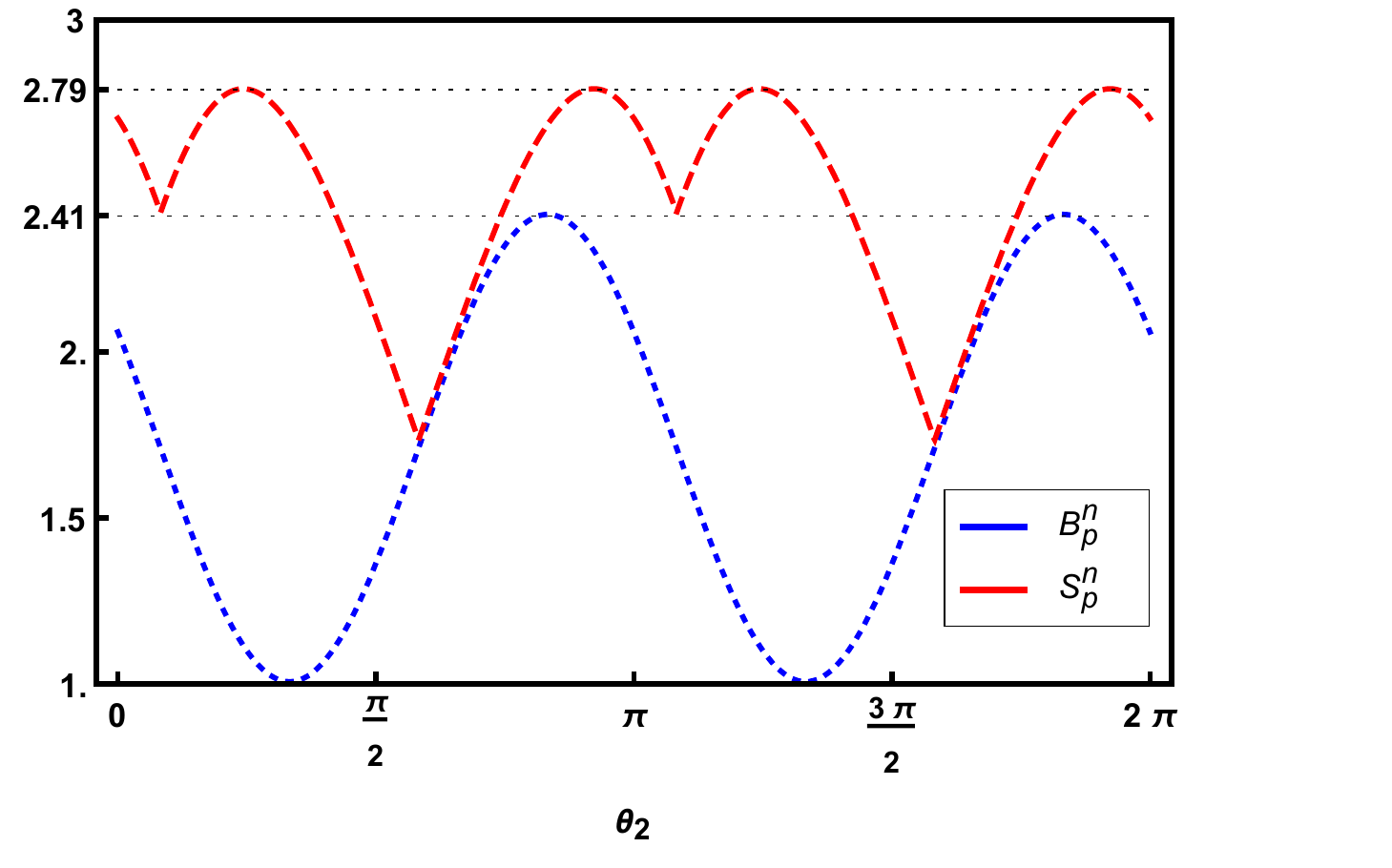}
	\caption{Bell inequality violation versus steering inequality violation for N00N states. In the figure, we plot  $S^n_{p}$ and $|B^n_{p}| $  with respect to $\theta_2$ at $\phi_1 = -0.13$, $\phi_2 = 0.65$, and $\theta_1=0.26$.}
	\label{fig:1}
\end{figure}
 In Fig. 3, we have plotted $S^n_{p}$ and $|B^n_{p}| $ with respect to $\theta_2$ with two particles in each system, and  for  $\phi_1 = -0.13$, $\phi_2 = 0.65$, and $\theta_1=0.26$. The profiles are the same as in Fig. 1, except for a scaling of the horizontal axis. The number of maxima for the steering inequality violation on the \(\theta_2\)-parameter space is four, while in the same space, the number of maxima is two for Bell inequality violation. 
 

\section{Summary}
We have checked for violation of a steering inequality for quantum states of systems of indistinguishable particles that are restricted by a 
%
particle-number-superselection rule.   In order to bypass the superselection rule, we have  formulated a steering inequality, where  measurements on two copies of the system state is considered, a strategy that had previously been applied for Bell inequality violation. We found that the restriction of  superselection rule on the measurement space prevents  violation of the steering inequality from reaching its maximal quantum value. Violation was checked for non-interacting Bose-Einstein condensate states, for which we also found that  as the particle number increases, violation of the steering inequality decreases. We also checked for violation in N00N states. For both the types of states,  steering inequality violations are higher than  Bell inequality violations, implying that the former will have a higher visibility against admixture of white noise, in certain situations.


\acknowledgements
We acknowledge partial support from the Department of Science and Technology, Government of India through the QuEST grant
(grant number DST/ICPS/QUST/Theme-3/2019/120).

\end{document}